\documentclass[fleqn,usenatbib]{mnras}

\usepackage[T1]{fontenc}
\usepackage{ae,aecompl}

\usepackage{graphicx}	
\usepackage{gensymb}
\usepackage{amsmath}	
\usepackage{amssymb}	

\newcommand{\be}{\begin{equation}}
\newcommand{\ee}{\end{equation}}
\newcommand{\bea}{\begin{eqnarray}}
\newcommand{\eea}{\end{eqnarray}}

\newcommand{\msun}{{\rm M}_\odot}

\newcommand{\ifm}[1]{\relax\ifmmode#1\else$\mathsurround=0pt #1$\fi}
\newcommand{\kms}{\ifmmode\,{\rm km}\,{\rm s}^{-1}\else km$\,$s$^{-1}$\fi}

\newcommand{\ltsima}{$\; \buildrel < \over \sim \;$}
\newcommand{\lsim}{\lower.5ex\hbox{\ltsima}}
\newcommand{\gtsima}{$\; \buildrel > \over \sim \;$}
\newcommand{\gsim}{\lower.5ex\hbox{\gtsima}}

\def\la{\langle}

\usepackage{color}

\newcommand{\Nsamp}{385}

\usepackage{soul}
\usepackage{ulem}
\usepackage{tabularx}
\usepackage{caption}
\usepackage[table,x11names]{xcolor}

\usepackage{newtxtext,newtxmath}

\title[Impact of gas flows on galaxy scaling relations]{MusE GAs FLOw and Wind (MEGAFLOW) IX. The impact of gas flows on the relations between the mass, star formation rate and metallicity of galaxies}

\author[Langan et al.]{Ivanna Langan$^{1,2}$,\thanks{E-mail: ivanna.langan@eso.org}
Johannes Zabl$^{2,3}$,
Nicolas F. Bouché$^{2}$,
Michele Ginolfi$^{1}$,
\newauthor
Gergö Popping$^{1}$,
Ilane Schroetter$^{4}$,
Martin Wendt$^{5,6}$,
Joop Schaye$^{10}$,
\newauthor
Leindert Boogaard$^{7,10}$,
Jonathan Freundlich$^{8}$,
Johan Richard$^{2}$,
Jorryt Matthee$^{9}$,
\newauthor
Wilfried Mercier$^{4}$,
Thierry Contini$^{4}$,
Yucheng Guo$^{2}$,
Maxime Cherrey$^{2}$
\\
$^{1}$European Southern Observatory, Karl-Schwarzschild-Str. 2, D-85748, Garching, Germany \\
$^{2}$Univ Lyon, Univ Lyon1, Ens de Lyon, CNRS, Centre de Recherche Astrophysique de Lyon (CRAL) UMR5574,\\
F-69230 Saint-Genis-Laval, France\\
$^{3}$Institute for Computational Astrophysics and Department of Astronomy \& Physics, Saint Mary’s University, 923 Robie Street,\\
Halifax, Nova Scotia, B3H 3C3, Canada \\
$^{4}$ Institut de Recherche en Astrophysique et Plan\'etologie (IRAP), Universit\'e de Toulouse, CNRS, UPS, F-31400 Toulouse, France\\
$^{5}$Institut f\"{u}r Physik und Astronomie, Universit\"{a}t Potsdam, Karl-Liebknecht-Str. 24/25, 14476 Golm, Germany \\
$^{6}$Leibniz-Institut f\"{u}r Astrophysik Potsdam, An der Sternwarte 16, D-14482 Potsdam, Germany\\
$^{7}$ Max Planck Institute for Astronomy, K\"{o}nigstuhl 17, 69117, Heidelberg, Germany\\
$^{8}$Universit\'{e} de Strasbourg, CNRS UMR 7550, Observatoire Astronomique de Strasbourg, F-67000 Strasbourg, France\\
$^{9}$Department of Physics, ETH Z\"{u}rich, Wolfgang-Pauli-Strasse 27, CH-8093 Z\"{u}rich, Switzerland\\
$^{10}$Leiden Observatory, Leiden University, PO Box 9513, 2300 RA Leiden, the Netherlands}

\date{Accepted XXX. Received YYY; in original form ZZZ}

\pubyear{2019}

\begin{document}
\begin{sloppypar}
\label{firstpage}
\pagerange{\pageref{firstpage}--\pageref{lastpage}}
\maketitle

\begin{abstract}
We study the link between gas flow events and key galaxy scaling relations: the relations between star formation rate (SFR) and stellar mass (the main sequence, MS), gas metallicity and stellar mass (the mass-metallicity relation, MZR) and gas metallicity, stellar mass and SFR (the fundamental metallicity relation, FMR). Using all star-forming galaxies (SFGs) in the 22 MUSE fields of the MusE GAs FLOw and Wind (MEGAFLOW) survey, we derive the MS, MZR and FMR scaling relations for \Nsamp\ SFGs
with $M_\star = 10^{8} - 10^{11.5}\,\msun$ at redshifts $0.35<z<0.85$. 
Using the MUSE data and complementary X-Shooter spectra at $0.85<z<1.4$, we determine the locations of 21 SFGs associated with inflowing or outflowing circumgalactic gas (i.e. with strong MgII absorption in background quasar spectra) relative to these scaling relations.  Compared to a control sample of galaxies  without gas flows (i.e., without MgII absorption within 70 kpc of the quasar), SFGs with inflow events (i.e., MgII absorption along the major axis) are preferentially located above the MS, while SFGs with ouflow events (i.e., \ion{Mg}{II} absorption along the minor axis) are preferentially more metal rich. Our observations support the scenario in which gas accretion increases the SFR while diluting the metal content and where circumgalactic outflows are found in more metal-rich galaxies.
\end{abstract}




\section{Introduction}
\label{sec:intro}

The flow of baryons in and out of galaxies is a fundamental component of galaxy formation and evolution, controlling the buildup of stars and metals in galaxies (for recent reviews see e.g., \citealt{Somerville2015}, \citealt{CGM2017}, \citealt{Peroux2020}, \citealt{Walter2020}).  Inflows of fresh material coming from the circumgalactic medium (CGM) are believed to drive star formation activity which increases the metallicity {($Z$, 12+log(O/H))} and stellar mass {($M_\star$)} of galaxies. Outflows, which are a consequence of internal processes such as supernovae or active galactic nuclei (AGN), expel the gas out of galaxies and can suppress inflows. By doing so, outflows lower galaxy gas content and prevent the formation of new stars. The cycle of baryons in and out of galaxies is thus at the core of galaxy formation and evolution (e.g., \citealt{Tacchella2016}, \citealt{Mitchell2022}).

This scenario is supported by observations and simulations that have shown that several major galaxy properties correlate, forming so-called scaling relations. Amongst the most notable correlations, those of interest for this work are: (i) the correlation between $M_\star$ and SFR, dubbed the ``main sequence'' (MS; \citealt{Noeske}, \citealt{Elbaz}, \citealt{Speagle2014}, \citealt{Whitaker2014}, \citealt{Renzini2015}, \citealt{Boogaard2018}) and (ii) the correlation between $M_\star$ and gas metallicity, referred to as the ``mass-metallicity relation'' (MZR; \citealt{Tremonti04}, \citealt{Maiolino2008}, \citealt{Erb}, \citealt{Zahid}, \citealt{Sanders2022}). The MS shows that $M_\star$ and SFR are positively correlated, meaning that an increasing stellar mass is directly linked to an increasing SFR, over $10^7$ M$_\odot$ $\leq M_\star \leq 10^{11.5}$ M$_\odot$. This relation has been observed to hold up to redshifts $z=4-5$ with decreasing data in the lower mass range as it becomes increasingly difficult to observe faint galaxies at high redshifts (\citealt{Speagle2014}, \citealt{Boogaard2018}). The mass-metallicity relation tells us that star-forming galaxies (SFGs) with a high stellar mass also have a high gas metallicity, at least for galaxies with $10^{7.5}$ M$_\odot$ $\leq M_\star \leq 10^{11.5}$ M$_\odot$. This has been observed up to redshift $z=3.5$ \citep{Maiolino2019}. At masses larger than $M_\star=10^{11.5}$ M$_\odot$, the relation flattens and reaches a metallicity plateau. 


These scaling relations can be interpreted as the reflection of an equilibrium between galactic inflows, star formation, and outflows. This equilibrium is often referred to as the ``bathtub model'' \citep{Bouche2010} or ``gas-regulator'' \citep{Lilly}. Under this equilibrium context, the SFR is regulated by the accretion rate, which is itself dependent on the galaxy halo/stellar mass. Given that galaxies evolve on tracks that tend to be orthogonal to the MS below $z=3$ \citep{Bouche2010,Matthee2019} and that the metallicity tends to always increase with time (dZ/dt$>0$), the SFR-$M_\star$-time results in a SFR-$M_\star$-$Z$ relation, which is akin to the MZR.

The scatter around the MS and the MZR has generated interest in the last decade as it provides precious insights into different physical processes that galaxies undergo \citep[e.g.][]{Mitra2017,Tacchella2016,Torrey2018,Matthee2019}. One can expect to find galaxies to lie on the upper part of the MS, likely due to recent events of gas accretion, while galaxies on the lower part of the MS typically have lower gas fractions \citep{Saintonge2016}, likely driven by a combination of gas depletion, outflows driven by recent active star-formation and suppression of inflows. Hence, individual galaxies may
continuously cycle through occupying the upper and lower envelope of the main sequence partly in response to gas-accretion and gas outflows over `short' timescales of a few Gyr ($\sim1-3$Gyr at $z\sim0.6$, \citealt{Tacchella2016,Torrey2018}).
Alternatively, the position of galaxies relative to the main sequence may simply reflect the scatter in formation times at fixed mass \citep{Gladders2013,Abramson2016}. Cosmological simulations, in particular EAGLE \citep{Schaye2015}, have shown that while both short and long term fluctuations do occur, the scatter in the main sequence is actually dominated by fluctuations occurring on very long timescales ($\gg 1$ Gyr), driven by the time at which the halo formed \citep{Matthee2019}.

The scatter around the MZR has been shown to correlate with SFR, resulting in the so-called ``fundamental metallicity relation" between SFR, metallicity and stellar mass (FMR; \citealt{Mannucci2010}, \citealt{LaraLopez2010}, \citealt{Cresci2012}, \citealt{Andrews2013}). This FMR agrees qualitatively with various analytical/numerical models which have attempted to account for the scatter of the MZR with regards to SFR  \citep[e.g.][]{Forbes2014,Torrey2019,vanLoon2021,Wang2021}. These models have shown that the MZR scatter can be attributed to the competing impact of gas accretion and star formation.
As fresh metal-poor gas is being accreted onto a galaxy, it feeds the gas reservoir resulting in an increase of SFR while at the same time diluting the gas-phase metallicity (e.g., \citealt{Dave2011}, \citealt{Somerville2015}, \citealt{Torrey2018}, \citealt{Matthee2018}, \citealt{vanLoon2021}). Consequently, the gas-phase metallicity decreases with an increasing SFR at fixed stellar mass (e.g., \citealt{Mannucci2010}, \citealt{LaraLopez2010}, \citealt{Lagos2016}, \citealt{Derossi2017}). 

Thus, it is clear that gas flows play a relevant role in the formation and evolution of galaxies. However, measuring gas flows and their properties is a difficult task to achieve. Due to a scarcity of measured inflow and outflow rates for statistical samples of galaxies, observational studies linking the flow rate of gas to galaxy stellar mass, metallicity and SFR are lacking. Until today, there are hardly any direct observational constraints regarding gas flows and how they correlate to scaling relations, except perhaps with respect to the main sequence and outflows (\citealt{ForsterSchreiber2019}, \citealt{RobertsBorsani2019}, \citealt{Avery2021}). In recent years, techniques have been developed aiming to measure gas flows in the CGM using quasar absorption studies. This technique can play a crucial role here, as it has the power to effectively identify the presence of in- and outflows for statistical samples of galaxies over time, thanks to absorption in the spectra of background sources (e.g., \citealt{paperI}, \citealt{Ho2017}, \citealt{paperII}, \citealt{paperIII}).

In this paper, we study the impact of gas flows on galaxy properties. In particular, how gas flows are linked to the main-sequence and the fundamental metallicity relation. This paper is organised as follows. In Section~\ref{sec:data}, we describe the sample this study is based on. Section~\ref{sec:methods} presents the methods we use to carry out the analysis. In Section~\ref{sec:results}, we present the results, i.e our analysis of the main-sequence and the fundamental metallicity relation. We discuss our results in Section~\ref{sec:discussion} and summarise our findings and conclusions in Section~\ref{sec:summary}. The cosmology assumed throughout this work follows the $\Lambda$CDM standard cosmological parameters: $H_0 = 70 $km~s$^{-1}$Mpc$^{-1}$, $\Omega_m = 0.3$ and $\Omega_\Lambda = 0.7$. Throughout this work, we use a \citet{Chabrier2003} stellar initial mass function (IMF).


\section{Data: MUSE AND XShooter}
\label{sec:data}

\subsection{MUSE data from the MEGAFLOW survey}
\label{subsec:survey}

The MUSE GAs FLOw and Wind (MEGAFLOW) survey was designed to study the properties of gas flows surrounding star-forming galaxies (SFGs) via background quasars using the Multi Unit Spectroscopic Explorer (MUSE, \citealt{Bacon2010}) spectrograph on the Very Large Telescope (VLT). The survey consists of 22 quasar fields selected with {\it multiple} \ion{Mg}{II} absorption lines (rest equivalent width $W_r^{\lambda2796} > 0.5 - 0.8$\AA), from the \citet{Zhu} catalogue based on the Sloan Digital Sky Survey (SDSS; \citealt{Ross}; \citealt{Alam}). This criterion resulted in 79 strong \ion{Mg}{II}  absorbers over the 22 quasar spectra.
For each quasar, high-resolution spectroscopic follow-up observations with the Ultraviolet and Visual Echelle Spectrograph (UVES, \citealt{Dekker2000}) were carried out. 

Several papers have already been published as part of the MEGAFLOW survey, we refer the reader to the following papers for a more detailed understanding of the observational strategy \citet{paperI} (hereafter paper I) and data reduction, in particular: \citet{paperII} (hereafter paper II) which focuses on accretion cases and \citet{paperIII} (hereafter paper III) which focuses on wind cases. 

\subsection{CGM gas flows}
\label{subsec:gasflows}

Among the galaxies in the 22 MUSE fields at the \ion{Mg}{ii} absorber redshift, we select those with an impact parameter ($b$, the distance between the galaxy and the apparent position of the background quasar) less than $b < 100$ kpc in order to ensure the gas we are probing likely belongs to a single galaxy. In addition, we select galaxies that have no major companion within 100~kpc and no signs of merger (see section 4.1 and 4.2 of paper~II for a more detailed explanation of the target selection strategy).

\begin{figure}
    \begin{center}
        \includegraphics[width=\columnwidth]{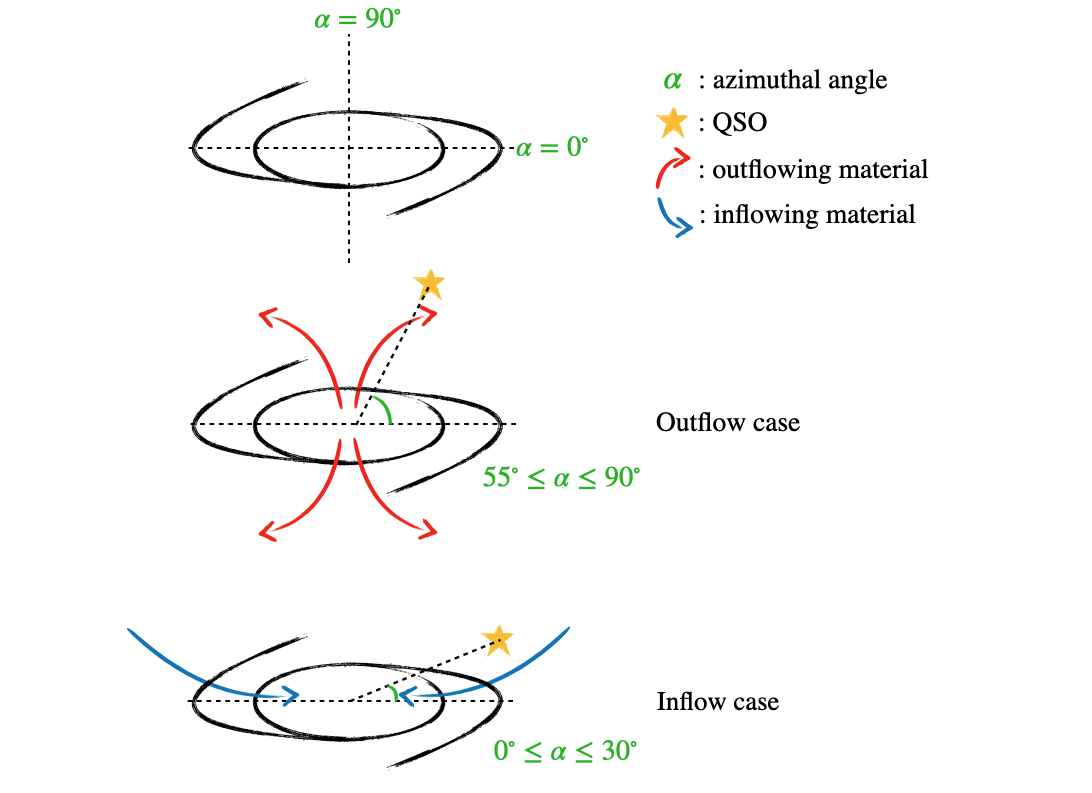}
        \caption{Schematic illustrating how galaxies are categorised as ``outflow'' or ``inflow'' cases, depending on the azimuthal angle $\alpha$ between the apparent QSO location and the galaxy major-axis. The black lines represent an edge-on galaxy. The red arrows pointing away from the galaxy  represent an outflow with material being ejected out of the galaxy. In the case of outflows, $\alpha \geq 55^{\circ}$. The blue arrows pointing towards the galaxy represent inflow material being accreted onto the galaxy. In the case of inflows, $\alpha \leq 30^{\circ}$.}
        \label{fig:outinflows}
    \end{center}
\end{figure}

Then, as in  paper~II (section 4.3), we classify the galaxy-quasar pairs according to the apparent location of the quasar 
with respect to the host galaxy motivated by the established bimodality in the azimuthal distribution of the \ion{Mg}{II} absorption around galaxies  \citep[][paper~II, paper~III, Guo et al., in prep.]{Bordoloi2011,Bouche2012, Kacprzak2012,Bordoloi2014,Lan2018,Lundgren2021}.  As in paper II/III, when the azimuthal angle, $\alpha$, defined as the angle between the apparent location of the background quasar and the major-axis of the galaxy~\footnote{The major-axis is determined by fitting a 3D morpho-kinematical model to the [OII] emission doublet using the 3D fitting tool \textsc{GalPack$^{3D}$} (\citealt{Galpak}).} associated (for an illustration, see Figure \ref{fig:outinflows}) is $55\degree \leq \alpha \leq 90\degree$, i.e. the quasar sightline is close to the galaxy’s minor axis ($\alpha = 90\degree$) and  likely probing outflows, we classify such case as an ``outflow''-selected galaxy. When the sightline passes close to the major axis ($0\degree \leq \alpha \leq 30\degree$) and likely passing through extended gaseous discs associated with inflows,  we classify these cases as an ``inflow''-selected galaxy. 
Hereafter, for simplicity, we call galaxies selected on their major (minor) axis as ``inflow" or (``outflow'') galaxies, respectively.

Naturally, one should keep in mind that an ``inflow/outflow'' galaxy might also be undergoing an outflow/inflow event. However, given the travel time between the absorber and the galaxy, the inflow and outflow events could be separated in time by 0.5-1 Gyr (see section \ref{subsection:timescales}). Hence, an inflow-selected galaxy might not have had the time to develop a strong outflow event associated with this inflow and similarly an outflow-selected galaxy might no longer be related to a strong accretion event.

\subsection{X-Shooter data}
\label{subsec:xshooter}
In addition to the MUSE data, we observed 13 galaxies with identified outflows/inflows from the MEGAFLOW survey, with the VLT multi-wavelength spectrograph X-Shooter \citep{Xshooter} over 20.2h (see Zabl et al. in prep). One of the objectives of the proposal associated to this data set was to measure rest-optical strong emission lines, namely H$\beta$ ($\lambda_{\text{rest}} = 4861$ \AA), [{\rm O}{\sc \,ii}]  ($\lambda_{\text{rest}} = 3727,3729$ \AA), [{\rm O}{\sc \,iii}] ($\lambda_{\text{rest}} = 4959,5007$ \AA), H$\alpha$ ($\lambda_{\text{rest}} = 6564$ \AA) and [{\rm N}{\sc \,ii}] ($\lambda_{\text{rest}} = 6585$ \AA), to measure gas-phase metallicities. The data was reduced following the X-Shooter pipeline version 3.5.0 (\citealt{Modigliani2010}, \citealt{Sparre2015}) with a small modification for cosmic rays. A more detailed description of the data reduction steps can be found in \citet{Zabl2015} and \citet{Man2021}.




\subsection{Sample selection and characteristics}
\label{subsec:sample}

From the MEGAFLOW full survey (Bouché et al. in prep.), there are $\sim$1000 [\ion{O}{ii}]
emitters with a S/N ratio greater than 3. Using mock [{\rm O}{\sc \,ii}] emitters with realistic kinematic properties and sizes and taking into account the MUSE sensitivity limit (Bouché et al. in prep), we found a $50\%$ completeness level at $\approx 8 \times 10^{-18} \text{erg}\,\text{s}^{-1}\text{cm}^{-2}$. This translates into a $\text{SFR}$ limit of $\approx 0.08 \, \text{M}_\odot\,\text{yr}^{-1}$ at $z=0.5$, the median redshift of our sample.

In order to have galaxies with metallicity diagnostic lines such as [{\rm O}{\sc \,iii}], $H\beta$ and [{\rm O}{\sc \,ii}] fall in the MUSE wavelength range, we selected galaxies in the redshift range $z=0.35$ to $z=0.85$.  To ensure a homogeneous sample of galaxies with metallicities from all tracers based on all the  lines, i.e., [{\rm O}{\sc \,iii}], $H\beta$ and [{\rm O}{\sc \,ii}] (see Section \ref{subsec:metallicitytracers} hereafter), we only keep galaxies for which all the three lines have, simultaneously,  a signal-to-noise ratio (SNR) higher than 1.
There are 385 galaxies with [{\rm O}{\sc \,ii}], $H\beta$ and [{\rm O}{\sc \,ii}] lines at $z = 0.35-0.85$. These have stellar masses ranging from $M_\star = 10^{6.2}$ M$_\odot$ to $M_\star = 10^{11.5}$ M$_\odot$. Assuming a Chabrier IMF, the stellar masses are estimated using a SED fitting code using the stellar continuum with a delayed-$\tau$ star-formation history and dust extinction law following \citep{Calzetti2000} (\textit{coniecto}; see section 5.6.1 of \citealt{Zabl2016}). This sample of 385 galaxies constitutes the basis for our study.

%

%

Of these 385 galaxies with MUSE metallicities (section~{3.2}), there are 15  galaxies with known CGM outflows or inflows (2 ``inflow'' galaxies from paper II and 13 ``outflow'' galaxies from paper III). From the 13 MEGAFLOW galaxies with X-Shooter data, there are  6 additional galaxies at $0.91<z<1.42$ with known in/outflows from paper II/III (5 ``inflow'' galaxies and 1 ``outflow'' galaxy) and 7 galaxies are in common to the 15 with known gas flows with MUSE metallicities. We use these 7 galaxies in common to check the validity of only using redshifted to the NIR strong emission lines to derive the metallicity (see section \ref{subsec:metallicitytracers} and table \ref{tab:xshooter_muse_comp} for the validity check).


In total, we have a sample of 21 ``inflow/outflow'' galaxies with metallicity indicators, 14 cases of outflow galaxies and 7 cases of inflow galaxies, with impact parameters ranging from, respectively, $9.6$ kpc to $69.0$ kpc and $17$ kpc to $65$ kpc. We list the galaxy properties in Table \ref{tab:gasflowcases}.

\begin{table*}
    \centering
    \caption{Summary of ``inflow/outflow'' galaxies properties with metallicity measurements used in this study.  (1) Field ID; (2) Galaxy ID from the MEGAFLOW catalog (Bouché et al. in prep.); (3) Galaxy redshift; (4) Galaxy stellar mass [$\log_{\text{10}}$(M$_\odot$)] from SED fitting, with the exception of galaxies 112005 and 272002 for which the stellar mass is the average between the dynamical mass and the stellar mass derived from the SED fit (due to a non-negligible difference between the two stellar masses); (5) Star formation rate [M$_\odot$\,yr$^{-1}$]; (6) Gas-phase metallicity [12+$\log(O/H)$]; (7) Mass accretion rate (negative value) or mass outflow rate (positive value) [M$_\odot$\,yr$^{-1}$]; (8) Impact parameter [kpc]; (9) W = wind case, A = accretion case; (10) Metallicity indicators from M = MUSE data, XS = X-Shooter data or MXS = both; (11) Reference from paper II or III.}
    \begin{tabularx}{2\columnwidth}{ccccccrXXXX}
	    \hline
	    Field & ID  & $z$ & $M_\star$ & SFR &  $12+\log(O/H)$ & $\dot{M}_{\text{in/out}}$ & $b$ & Type & I & P\\
	    (1) & (2) & (3) & (4) & (5) & (6) & (7) & (8) & (9) & (10) & (11)\\
	    \hline
	    \hline
	    J0014m0028 & 11100 & 0.83389 & 8.35 & 1.47 & 8.52 & +3.3 & 9.70 & W & M & III\\
	    
	    J0015m0751 & 13063 & 0.73143 & 10.38 & 2.07  & 8.70 & +5.3 & 35.5 & W & MXS & III\\
	    
	    J0015m0751 & 13128 & 0.50814 & 9.99 & 1.35  & 8.63 & +4.2 & 24.1 & W & MXS & III\\
	    
	    J0015m0751 & 13085 & 0.81654 & 10.13 & 2.48  & 8.74 & +9.2 & 20.7 & W & MXS & III\\
	    
	    J0103p1332  & 15040 & 0.78817 & 9.42 & 2.68 & 8.47 & -0.9 & 20.0 & A & MXS & II\\
	    
	    J0131p1303 & 16070 & 1.01040 & 8.80 & 6.85  & 8.39 & +1.6 & 26.4 & W & XS & II\\
	    
	    J0145p1056 & 18049 & 0.54994 & 9.36 & 0.39  & 8.71 & -0.7 & 22.0 & A & M & II\\
	    
	    J0145p1056 & 18063 & 0.76993 & 9.60 & 0.52 & 8.84 & +1.0 & 12.9 & W & M & III\\
	    
	    J0800p1849 & 19105 & 0.6082 & 9.50 & 4.01 & 8.44 & -1.8 & 65.0 & A & MXS & II\\
	    
	    J0800p1849 & 19073 & 0.84331 & 9.92 & 2.75 & 8.58 & +1.0 & 20.9 & W & MXS & III\\
	    
	    J0937p0656 & 21109 & 0.70185 & 9.24 & 2.23 & 8.49 & +1.2 & 69.0 & W & MXS & III\\
	    
	    J0937p0656 & 21113 & 0.70191 & 9.87 & 4.87 & 8.60 & +0.3 & 38.7 & W & MXS & III\\
	    
	    J1039p0714 & 22066 & 0.81849 & 10.19 & 1.89 & 8.71 & +21.2 & 24.5 & W & M & III\\
	    
	    J1039p0714 & 22056 & 0.94927 & 9.80 & 7.69 & 8.64 & -2.7 & 49.0 & A & XS & II\\
	    
	    J1236p0725 & 25067 & 0.63817 & 10.35 & 4.94 & 8.75 & +1.0 & 66.9 & W & M & III\\
	    
	    J1236p0725 & 25095 & 0.91280 & 10.50 & 33.62 & 8.49 & -12.5 & 17.0 & A & XS & II\\
	    
	    J1352p0614 & 27052 & 0.60420 & 8.85 & 0.04 & 8.65 & +1.1 & 14.0 & W & M & III\\
	    
	    J1358p1145 & 28056 & 0.80923 & 9.61 & 4.26 & 8.58 & +4.2 & 12.7 & W & MXS & III\\
	    
	    J1358p1145 & 28065 & 1.41713 & 9.80 & 19.90 & 8.45 & -17.7 & 30.0 & A & XS & II\\
	    
	    J1425p1209 & 29002 & 0.59678 & 8.70 & 2.09 & 8.33 & +0.7 & 9.60 & W & M & III\\
	    
	    J2152p0625 & 32055 & 1.05287 & 10.15 & 11.49 & 8.47 & -1.0 & 49.0 & A & XS & II \\
	    \hline
    \end{tabularx}
    \label{tab:gasflowcases}
\end{table*}


\section{Methods}
\label{sec:methods}

\subsection{Flux measurements}
\label{subsec:fluxmeasurements}
For the galaxies with MUSE data, we measure their flux from their spectra using \texttt{pyplatefit}, which is an improved Python-based version of the PLATEFIT IDL software developed by \citet{Brinchmann2004}. \texttt{pyplatefit} fits the continuum and emission spectra simultaneously. We apply a global fitting of all lines at a given redshift \citep[as in][]{Bacon2022} 
The cases in which one or more lines have relatively low SNR result in large errors on the metallicity tracers (see Figure \ref{fig:metallicitytracers} in Section \ref{subsec:metallicitytracers}). In figure \ref{fig:muselines} we show an example of the [{\rm O}{\sc \,iii}], $H\beta$ and [{\rm O}{\sc \,ii}] fits.

For the galaxies with X-Shooter data, we fit a gaussian to the available lines (or a double gaussian in the case of doublets) using \texttt{curvefit} from the SciPy library \citep{Scipy}. The availability of the lines depends on the quality of the data and redshift. In 2 out of the 6 X-Shooter only galaxies (see second paragraph of section \ref{subsec:sample}), the quality of the data from the visible arm of X-Shooter is high enough to measure fluxes from [{\rm O}{\sc \,iii}], $H\beta$ and [{\rm O}{\sc \,ii}], in addition to the  lines redshifted to the NIR ([{\rm N}{\sc \,ii}] and $H\alpha$). In all cases, we measure the flux from [{\rm N}{\sc \,ii}] and $H\alpha$. As [{\rm N}{\sc \,ii}] is a particularly faint line, we constrain the bounds of the fit by fixing the width of the [{\rm N}{\sc \,ii}] line to the width the $H\alpha$ line. Then we check if the [{\rm N}{\sc \,ii}] line fitting gives a significant detection above the noise level. In case of a $< 3 \sigma$ detection, we set the [{\rm N}{\sc \,ii}] flux to be a $3 \sigma$ upper limit. Only one of the galaxies has a robust [{\rm N}{\sc \,ii}] detection (galaxy 52, see Table \ref{tab:gasflowcases}, with a $7.2\sigma$ detection). For the other 5 galaxies we set a $3 \sigma$ upper limit. In Figure \ref{fig:detanduplim} we show the robust [{\rm N}{\sc \,ii}] detection and an example of a $3 \sigma$ upper limit.


\subsection{Metallicity tracers}
\label{subsec:metallicitytracers}

In order to measure the gas-phase metallicity, we use the strong line method which uses strong nebular emission lines to derive the metallicity (see \citealt{Maiolino2019} for a review). 
We simultaneously fit different metallicity tracers from different strong nebular emission lines, namely $R_3$ ([{\rm O}{\sc \,iii}] $\lambda 5007$/$H\beta$), $R_{23}$ ([{\rm O}{\sc \,ii}] $\lambda 3727,29$/$H\beta$ + [{\rm O}{\sc \,iii}] $\lambda 4959, 5007$/$H\beta$), $O_{32}$ ([{\rm O}{\sc \,iii}] $\lambda 5007$/[{\rm O}{\sc \,ii}]) and $N_2$ ([{\rm N}{\sc \,ii}]/$H\alpha$). Our choice of metallicity tracers, i.e line ratios, is based on the wavelengths covered by MUSE and X-Shooter. Using different metallicity tracers at once enables us to break potential degeneracies and infer more robust metallicity measurements (in cases of double-branched metallicity tracer, dust reddening, dependence on the ionization parameter).


To go from the different line ratios to metallicity, we adopt the calibration presented in \citet{Curti2020}. This calibration is carried out using a direct method based on electron temperature $T_e$ over the entire metallicity range spanned by SDSS galaxies. They define each metallicity tracer as a function of metallicity following polynomial relations $\log(R) = \sum_N c_n x^n$, where R is the considered metallicity tracer and $x$ is the oxygen abundance, i.e metallicity, normalised to the solar value ($Z_\odot$ = 8.69, \citealt{Asplund2009}). Table \ref{tab:ratio} summarises all information necessary to use the \citet{Curti2020} calibration with the set of metallicity tracers described above.
Furthermore, because the calibration from \citet{Curti2020} is made for intrinsic fluxes, we correct our flux measurements for dust  extinction. We did so by using the \citet{Cardelli1989} extinction law with
$f_\text{obs} = f_\text{int} \times 10^{-0.4\, k_\text{Cardelli} \, \text{E(B-V)}}$, where $f_\text{obs}$ is the measured flux,  $f_\text{int}$ is the intrinsic flux, $k_\text{Cardelli}$ is the extinction for a given wavelength for each line and E(B-V) is the colour excess coefficient. We estimated E(B-V) from the empirical relation between stellar mass and dust extinction based on Balmer decrement measurements in \citet{Gilbank2010}.

In the end, we determine the best-fit parameter, i.e. the gas-phase metallicity $Z$, given the data (flux measurements) and a model (\citealt{Curti2020} calibration). We have up to four metallicity tracers and one unknown variable. This can be solved using the least square method.

\small
\begin{equation}
    \chi^2(Z) = \sum_{i=1}^{n} \left(\frac{\log[R_{{\rm int},i}(\text{E(B-V)})]) - \log[R_{\rm model,i}(Z)]}{\sigma_i}\right)^2,
    \label{eq:chisquare}
\end{equation}
\normalsize
 where $n$ is the number of available metallicity tracers, $\log(R_{{\rm int},i})$ are the ratios from the flux measurements corrected for dust extinction,
$\log(R_{{\rm model},i})$ are the calibrated ratios from \cite{Curti2020} and $\sigma_i$ are the errors on the measured ratios.
Finding the minimum in Equation \ref{eq:chisquare} gives us for each galaxy a value for its metallicity $Z$, as well as the associated errors. For 4 galaxies in our sample of 6 galaxies with X-Shooter data, we only use $N_2$ ([{\rm N}{\sc \,ii}]/$H\alpha$) to derive the metallicity, due to poor data in the visible arm of X-Shooter. However, owing to the galaxies with X-Shooter data that overlap with the MEGAFLOW-based sample of outflow/inflow galaxies, we were able to check the validity of only using N2 to derive the metallicity. For these galaxies, we compared the metallicity derived from the three visible tracers ($R_3$, $R_{23}$, $O_{32}$) with the metallicity derived from $N_2$ only and found very good agreement within the errors (see Table \ref{tab:xshooter_muse_comp} in the appendix for the comparison).

Figure \ref{fig:metallicitytracers} shows the result of the dust correction as well as the calibration of the metallicity tracers for the galaxies from the MEGAFLOW sample. The x-axis shows the optimised solution to Equation \ref{eq:chisquare}, i.e., the metallicity, against the different metallicity tracers. In green are the data points representing the ``raw" measurements, i.e., the flux ratios that are not corrected for dust extinction. In blue we show flux ratios, the metallicity tracers, corrected for dust extinction. The orange line corresponds to the \cite{Curti2020} model. The impact of dust extinction depends on the wavelength differences of the contributing lines, which is indeed what we observe for $R_3$ where the dust correction does not impact the tracer much. As the wavelengths of [{\rm O}{\sc \,iii}] and H$\beta$ are close, their dust correction cancels out in the ratio. But for the other two tracers, $R_{23}$ and $O_{32}$, we can see that the dust correction is non-negligible.


\begin{table}
	\centering
	\caption{Definition of metallicity tracers used for this study and their associated polynomial function coefficients from \citet{Curti2020}, with $\log(R) = \sum_N c_n x^n$.}
	\begin{tabular}{cccccc}
		\hline
		Line ratio & $c_0$ & $c_1$ & $c_2$ & $c_3$ & $c_4$ \\
		\hline
		\hline
		 \\
		$R_3 = \frac{[\text{O III}]}{\text{H}\beta}$ & -0.277 & -3.549 & -3.593 & -0.981 \\
		 \\
		\hline
		 \\
		$R_{23} = \frac{[\text{O II}]+[\text{O III}]}{\text{H}\beta}$ & 0.527 & -1.569 & -1.652 & -0.421 \\
		 \\
		\hline
		 \\
		$O_{32} = \frac{[\text{O III}]}{[\text{O II}]}$ & -0.691 & -2.944 & -1.308 \\
		\\
        \hline
        \\
		$N_2 = \frac{[\text{N II}]}{H\alpha}$ & -0.489 & 1.513 & -2.554 & -5.293 & -2.867\\
		\\
		\hline
	\end{tabular}
	\label{tab:ratio}
\end{table}

\begin{figure*}
    \begin{center}
        \includegraphics[width=2\columnwidth]{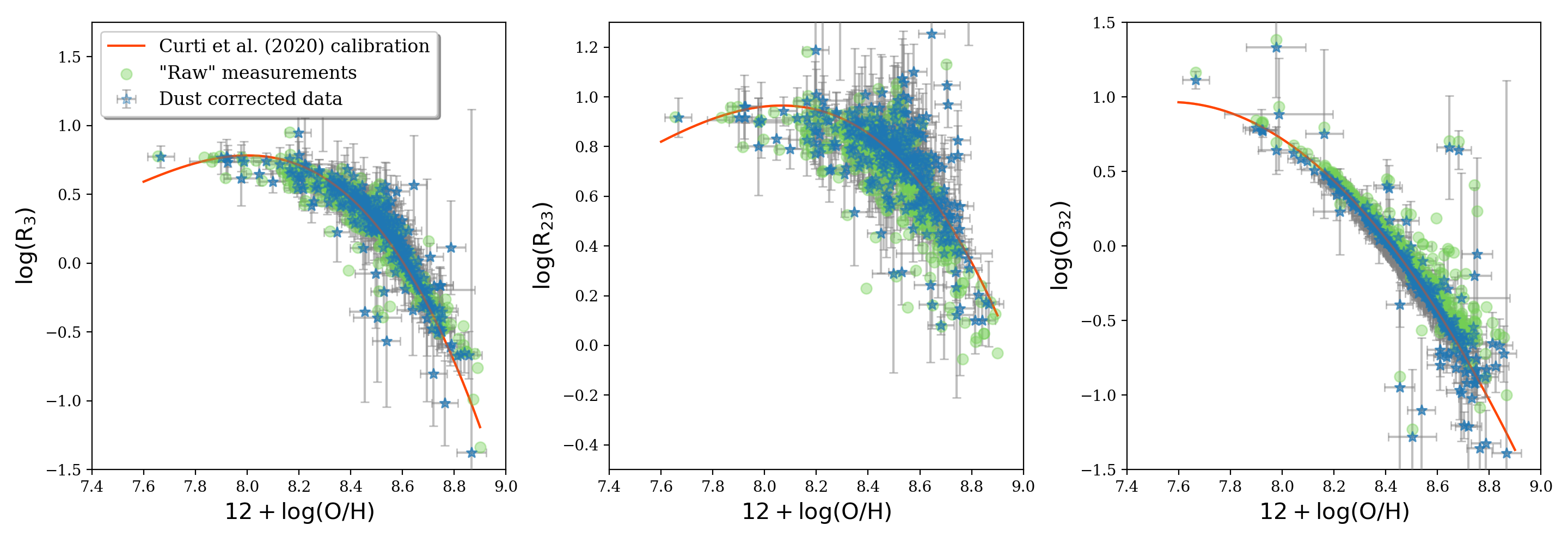}
        \caption{Metallicity tracers used in this study, i.e line ratios, as functions of the optimised metallicity according to Eq. \ref{eq:chisquare}. The left panel shows $R_3 = [\text{O III}]/\text{H}\beta$, the middle panel shows $R_{23} = ([\text{O II}]+[\text{O III}])/\text{H}\beta$ and the right panel shows $O_{32} = [\text{O III}]/[\text{O II}]$. For all three panels we show the following. "Raw" measurements, i.e., ratios using the line fitting results without any correction, are highlighted with green dots. Dust-corrected measurements are shown with blue stars with their associated error bars in grey. The corrected measurements are derived following \citet{Curti2020} calibration, which is represented with the orange line.}
        \label{fig:metallicitytracers}
    \end{center}
\end{figure*}



\subsection{Star formation rate calculation}
\label{subsec:sfr}

To determine the SFR for the galaxies in our MUSE sample (section \ref{subsec:sample}), we use the prescription from \citet{Gilbank2010}, who determined an empirical calibration of the SFR using the [{\rm O}{\sc \,ii}] line luminosity and the stellar mass, including the Balmer decrement for dust correction. Thus, we use the following.

\begin{equation}
    SFR = \frac{\text{L([{\rm O}{\sc \,ii}]})/(1.12\times2.53\times10^{40} {\rm erg \,s}^{-1})}{a \times \tanh[(x-b)/c]+d} ,
    \label{eq:SFROII}
\end{equation}
where L([{\rm O}{\sc \,ii}]) is the luminosity of the [{\rm O}{\sc \,ii}] line in erg s$^{-1}$, $x=\log(M_\star/$M$_\odot)$, $a=-1.424$, $b=9.827$, $c=0.572$ and $d=1.700$. We included the factor $1.12$ to convert the \citet{Gilbank2010} calibration, which is originally written for a \citet{Kroupa2001} IMF, to the \citet{Chabrier2003} IMF adopted in this work. 

As for our sample of 6 galaxies with X-Shooter data with $z>0.85$ (section \ref{subsec:sample}), we use the H$_\alpha$ line to derive the SFR. From the same \cite{Gilbank2010} prescription, we have:

\begin{equation}
    \text{SFR} = \frac{10^{0.4 A_{\text{H}_\alpha}}}{1.12} \frac{L(\text{H}_\alpha)}{1.27 \times 10^{41} {\rm erg \, s}^{-1}} ,
    \label{eq:SFRHalpha}
\end{equation}
where L(H$_\alpha$) is the observed luminosity of the H$_\alpha$ line in erg s$^{-1}$ and the dust extinction based on the Balmer decrement is $A_{\text{H}_\alpha} = a + b\log(M_*/M_\odot) + c\log(M_*/M_\odot)^2$, with a = 51.2, b = -11.199 and c = 0.615. The conversion factor $1.12$ is included to convert from the \citet{Kroupa2001} IMF to the \citet{Chabrier2003}.


\section{Results}
\label{sec:results}


In this work we aim at studying gas flows with regards to scaling relations in a self-consistent way, meaning the method and calibration used to derive the properties of the galaxies for which we have out/in-flows measurements are the same as for the galaxies defining the scaling relations. 


\subsection{The Main Sequence relation (MS)}
\label{subsec:MS}

Figure~\ref{fig:MS} (left)
shows  SFR as a function of $M_\star$, colour coded by the redshift of the galaxies. We find a main sequence (MS), with SFR increasing with stellar masses between $\sim10^7$ M$_\odot$ to $10^{11}$ M$_\odot$ (with a completeness limit around $M_\star \simeq 10^9 {\rm M}_\odot$), without clear evidence of a flattening of the relation. We fit the $M_\star$-SFR-redshift relation with the following plane equation:
\begin{equation}
    \log(\text{SFR}) = a_{\text{MS}}\;\log\left(\frac{M_*}{M_0}\right) + b_{\text{MS}}\;\log\left(\frac{1+z}{1+z_0}\right) + c_{\text{MS}}
    \label{eq:MS}
\end{equation}
where we take $M_0 = 10^{9.0}$ M$_\odot$ and $z_0 = 0.65$ corresponding to the median values for our sample. Using orthogonal least squares, we find the best-fit parameters $a_{\text{MS}} = 0.75 \pm 0.03$, {\bf $b_{\text{MS}} = 4.08 \pm 0.56$} and $c_{\text{MS}} = -0.14 \pm 0.03$. These numbers are summarised in Table \ref{tab:fitcoeffs}. We note that the redshift dependence is steeper compared to e.g. \citet{Boogaard2018}, albeit with the large uncertainty, it is consistent. In addition, we highlight in Figure \ref{fig:MS} galaxies for which we have outflow measurements with triangles with red-pink edges and galaxies for which we have inflow measurements with squares with blue edges. Looking at the positions of these galaxies with regard to the main sequence, it appears that `inflow' galaxies are located on the upper part of the MS relation while `outflow' galaxies do not seem to occupy a preferred position.

In order to facilitate the comparison between galaxies for which we have in/out-flow measurements to the main sequence, we plot the residuals of the MS in Figure \ref{fig:MS} (right panel), i.e $\log(\text{SFR}) -  0.73\log(M_\star) - 4.08\log(1+z) + 0.28$. As in the left panel, red/pink triangles show galaxies with outflows and blue squares show galaxies with inflows. One sees that all but one ``inflow'' galaxies are located above the zero line of the residuals and therefore show higher SFRs.

Figure~\ref{fig:histograms_res} is shown with the grey envelope in the right panel of Figure \ref{fig:MS} and is close to results found in previous work (e.g., \citealt{Speagle2014}).

\subsection{The (fundamental) mass-metallicity relation}
\label{subsec:FMR}

In Figure \ref{fig:FMR}, we show the FMR for the parent sample of 385 galaxies. The left-hand side of the Figure shows metallicity as a function of the $M_\star$, colour coded by the SFR of the galaxies. We find the well-known mass-metallicity relation (e.g., \citealt{Mannucci2010}, \citealt{Cresci2012}) with metallicity increasing linearly with stellar masses between $\sim10^7$ M$_\odot$ to $10^{11}$ M$_\odot$. Furthermore, we find the FMR where at a fixed stellar mass, higher metallicity galaxies have on average lower SFRs.

Similarly as in Figure~\ref{fig:MS}, we indicate in Figure~\ref{fig:FMR} galaxies selected in outflows (inflows)  with the symbols with red/pink (blue) edges. `Inflow' galaxies populate the lower part of the mass-metallicity relation (left panel), which also corresponds to higher SFRs in agreement with the main-sequence results (section \ref{subsec:MS}). ``Outflow'' galaxies, above our completeness limit of $10^9$ M$_\odot$, tend to be more metal rich at a given stellar mass.

We fit the $M_\star$-SFR-metallicity relation with the following plane equation:
\begin{equation}
    12+\log(O/H) = a_{\text{FMR}}\;\log\left(\frac{M_\star}{M_0}\right) + b_{\text{FMR}}\;\log\left(\frac{\rm SFR}{\rm SFR_0}\right) + c_{\text{FMR}}
    \label{eq:FMR_model}
\end{equation}
where again, we take $M_0 = 10^{9.0}$ M$_\odot$ and $\text{SFR}_0 = 0.40$ corresponding to the median values for our sample. Using orthogonal least squares, we find the best-fit~\footnote{The best-fit parameters are summarised in Table \ref{tab:fitcoeffs}.}:
\begin{equation}
    12+\log(O/H) = 0.22\;\log\left(\frac{M_\star}{M_0}\right) - 0.13\;\log\left(\frac{\rm SFR}{\rm SFR_0}\right) + 8.50
    \label{eq:FMR}
\end{equation}

In the right panel of Figure \ref{fig:FMR}, we show the residuals of the FMR defined as the difference between the predicted metallicity based on our best-fit from Eq.~\ref{eq:FMR} and the measured metallicity. From a Gaussian fit to these residuals, without taking into account the galaxies for which $\sigma(M_\star) > 0.4 \log(\text{M}_\odot)$ (grey circles), we find a scatter of $\approx0.1$ dex (right panel of Figure \ref{fig:histograms_res} in the appendix) which is shown with the grey envelop in the right panel of Figure \ref{fig:FMR}. This scatter is consistent with previous work (see \citealt{Maiolino2019} for a review).

As in Figure \ref{fig:MS}, pink triangles show ``outflow'' galaxies and blue squares show ``inflow'' galaxies. Outflow galaxies are predominantly located above the zero line, i.e. they are more metal rich than the parent sample, even after taking into account the SFR. However, inflow galaxies fall within the $2-\sigma$ scatter and do not occupy a specific region.

\subsection{Control sample}

In order to quantify how galaxies undergoing CGM outflow or inflow events behave with respect to galaxies without such a detected event, we built a control sample. The control sample is made of a subset of those galaxies among the parent sample of 385 MUSE galaxies (excluding the ``inflow/outflow'' galaxies) which have no detected outflows or inflows. Specifically, we selected galaxies with an impact parameter $b < 75$ kpc that have no absorber, i.e. a rest equivalent width (REW) of $W_r^{\lambda2796} < 0.1$ \AA,  within $250$ km/s of the galaxy redshift.

\begin{figure*}
    \begin{center}
        \includegraphics[width=2\columnwidth]{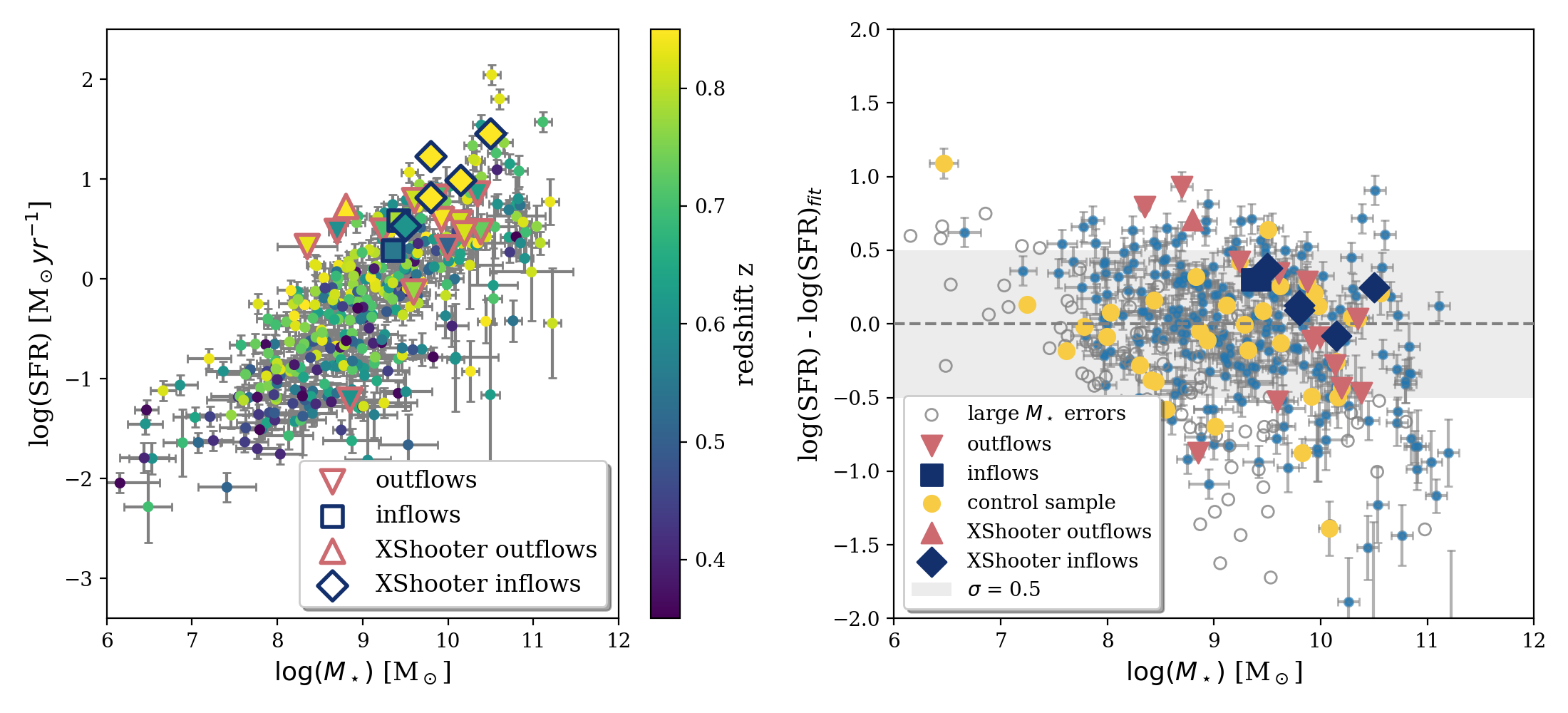}
        \caption{Left: The star formation rate of galaxies as a function of their stellar mass, namely the main sequence, colour coded with redshift. Right: Main sequence residuals around
        the SFR($M_\star,z)$ fit (grey dashed line). $68\%$ of the residuals are within the grey shaded area corresponding to  a scatter of $0.5$ dex. We highlight the control sample with yellow full dots. Galaxies for which the errors on the stellar mass are large, ie. $\sigma(M_\star) > 0.4 \log(\text{M}_\odot)$, are marked with grey circles. In both panels, galaxies for which we have CGM outflow measurements are shown with red triangles and galaxies for which we have CGM inflow measurements are shown with blue squares. ``Outflow'' galaxies do not occupy a specific location either on the MS plot and on the MS residuals plot. All but one ``inflow'' galaxies are located above the zero line of the residuals and therefore show higher SFRs.}
        \label{fig:MS}
    \end{center}
\end{figure*}

\begin{figure*}
    \begin{center}
        \includegraphics[width=2\columnwidth]{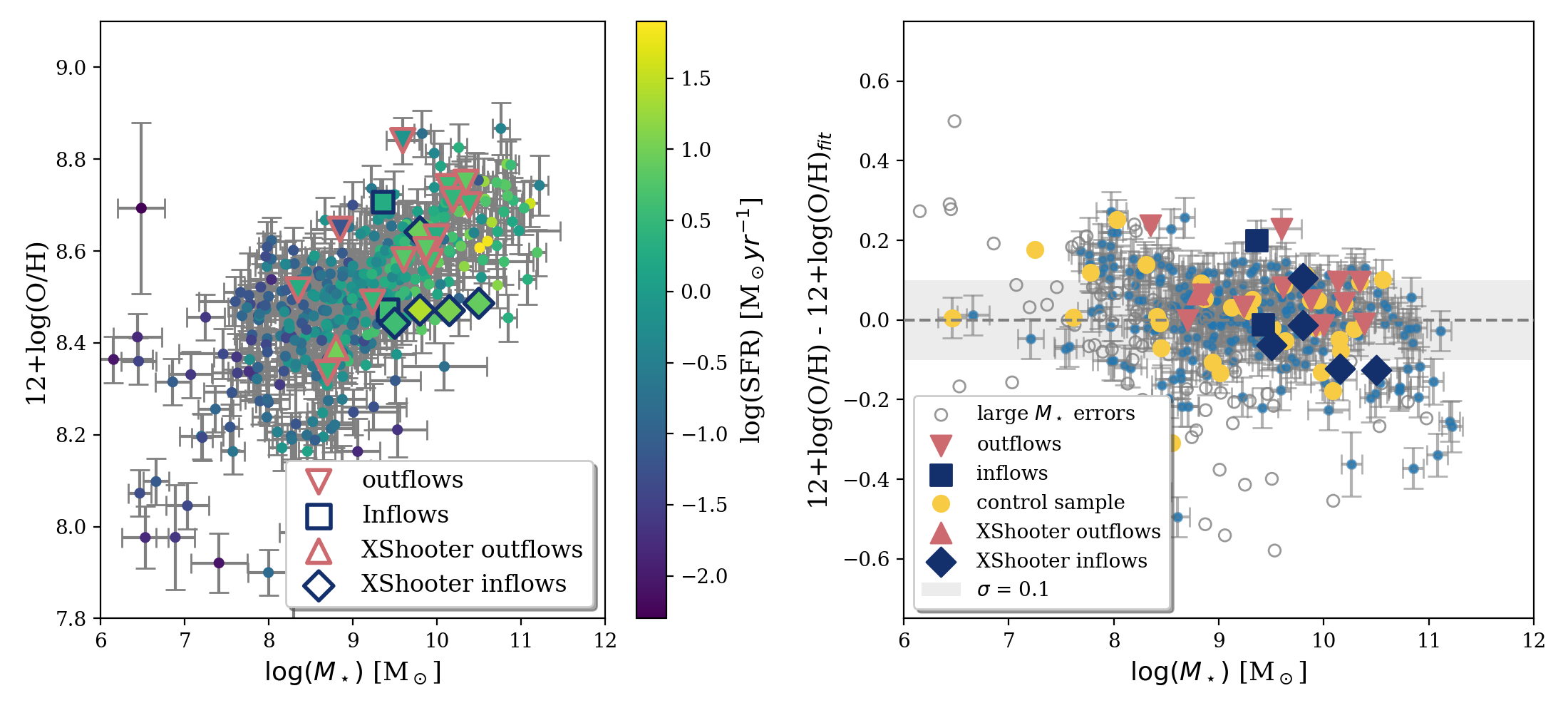}
        \caption{Left: The mass-metallicity relation (MZR), where the colour code shows the SFR. Right: The fundamental metallicity relation (FMR) residuals around the fit (grey dashed line). $68\%$ of the residuals are within the grey shaded area. We highlight the control sample with yellow full dots. Galaxies for which the errors on the stellar mass are large, i.e $\sigma(M_\star) > 0.4 \log(\text{M}_\odot)$, are marked with grey circles. In both panels, galaxies for which we have CGM outflow measurements are shown with red/pink triangles and galaxies for which we have CGM inflow measurements are shown with blue squares. Most of the ``inflow'' galaxies are located on the lower part of the MZR, so these galaxies have low metallicities and high SFRs (consistent with Figure \ref{fig:MS}). However, when taking into account the SFR dependence (FMR residuals), inflow galaxies don't show a specific location. On the other hand, all the ``outflow'' galaxies are on or above the zero-line of the FMR residuals, showing that these galaxies are more metal-rich.}
        \label{fig:FMR}
    \end{center}
\end{figure*}

This selection gives us a total of 35 galaxies, which are represented with filled yellow  circles in the residual plots of figures \ref{fig:MS} and \ref{fig:FMR} (right panels). We observe a homogeneous distribution of the galaxies belonging to the control sample around the zero line in the residual plots. We discuss the quantitative comparison between the outflow/inflow galaxies and this control sample in section \ref{sec:discussion}.

\begin{table}
	\centering
	\caption{Coefficients of the MS and FMR polynomial fits and their associated errors.}
	\begin{tabular}{c|crrccc}
	    \hline
		 & a & b & c & $\sigma_{a}$ & $\sigma_{b}$ & $\sigma_{c}$\\
		\hline
		\hline
		MS & 0.75 & 4.08 & -0.14 & 0.03 & 0.56 & 0.03\\
		\hline
		FMR & 0.22 & -0.13 & 8.50 & 0.02 & 0.02 & 0.01\\
		\hline
	\end{tabular}
	\label{tab:fitcoeffs}
\end{table}

\section{Discussion: the impact of gas flows}
\label{sec:discussion}

The main motivation for this paper is to study the potential impact of gas flows with respect to the galaxy's location on scaling relations such as the main sequence and the fundamental metallicity relation. These relations are typically interpreted as 
a reflection of the equilibrium between inflowing gas and outflowing gas in and out of galaxies in the context of the ``bathtub" model \citep{Bouche2010}, or ``regulator" model \citep{Lilly}. In the following we will discuss and interpret the results presented in Section \ref{sec:results} in the context of models and scenarios offered by theoretical works.

\subsection{The scatter around the main sequence}
\label{subsec:MSscatter}
Several groups have studied the MS and attempted to describe the origin of its scatter using numerical simulations \citep[e.g.][]{Tacchella2016,Lagos2016,Mitra2017,Torrey2018}.
Provided that SFGs react on a few dynamical times, the main results from these works are that SFGs form the MS as a reflection of the balance between gas accretion and outflow of gas, with galaxies lying on the upper part of the MS as a result of strong inflows and on the lower part as a result of outflows and suppression of inflows.
If instead SFGs fluctuate on the MS in several Gyrs
\citep{Matthee2019}, due to differences in  cosmological accretion driven by the halo formation time, then SFGs lying on the upper part of the MS are a result of stronger inflows.
We observe a tentative confirmation of these theoretical scenarios in our sample of galaxies associated to an inflow event, with all but one of these galaxies lying above the zero line of the MS residuals (Figure \ref{fig:MS}, right panel).
 
 To test quantitatively the tentative different distribution of the `inflow' galaxies, we used a KS-test to compare the distribution of the residuals of the inflow galaxies to the one of the control sample. We restrict ourselves to SFGs with $M_\star > 10^{9.0} M_\odot$, i.e., above our completeness limit, and find a p-value of 0.11. This p-value corresponds to a confidence level of 89\%.
Although with the sample available we cannot reject the null hypothesis that the distributions are the same at more than 1$\sigma$, the confirmation of the observed trend (5 out of 6 ``inflow'' galaxies above the MS) would require a larger samples.

Interestingly, we do not observe a preferential location for galaxies associated with an outflow event. This is perhaps not surprising given that SFGs above and below the MS are expected to have outflows. Depending on the star formation fluctuation time scales and on the wind travel time (see Section \ref{subsection:timescales}), one might expect stronger (in a cross-section sense) outflows for galaxies above or below the MS. The large scatter in loading factors 
 \citep{paperIII} prevents us from testing these scenarios.

 
\subsection{The scatter around the mass-metallicity relation}

Regarding the scatter around the MZR, according to numerical simulations (e.g., \citealt{Dave2017}, \citealt{Torrey2018}, \citealt{vanLoon2021}), galaxies found at the bottom part of the MZR show low metallicity but high SFR at a given stellar mass and this is typically attributed to recent metal-poor inflows that dilute the metal content of the galaxy while powering its SFR. Indeed, we observe most of our inflow galaxies at the bottom part of the MZR where the SFR is high and the metallicity low, at a given stellar mass. This observation also supports the scenario proposed by \citet{Zenocratti2022}, according to which disk galaxies form stars from the accretion of metal-poor gas and therefore show an anti-correlation between SFR and metallicity. In addition, we find that these inflow galaxies are distributed evenly around the zero line on the residual plot of the FMR (Figure \ref{fig:FMR}, right panel), suggesting that the dilution and increase of SFR happen almost simultaneously. This result supports the simulation
results of \citet{Torrey2018}
who found that the SFR and metallicity fluctuations are similar and that these fluctuations are opposite, a positive offset in SFR is associated to a negative offset in metallicity. \citet{Wang2021} also studied the correlations between SFR and metallicity on galactic scales and found two driving mechanisms. One being time-varying inflow rate and the other being time-varying star formation efficiency. They found a positive correlation in the latter case, while in the first case, i.e., time-varying inflow rate, the SFR and metallicity are negatively correlated. Therefore, our observations seem to support the time-varying inflow rate scenario.

\indent Furthermore, \citet{Torrey2019} and \citet{vanLoon2021} found that metallicity deviations tend to be anti-correlated with both outflow and inflow rates, i.e., the stronger the outflow/inflow, the lower the metallicity. This result does not seem to be confirmed in our observations. We actually observe that the inflow galaxies do not seem to have a preferential location on the FMR residuals plot (Figure \ref{fig:FMR}, right panel) but all the outflow galaxies are on or above the zero line in the FMR residuals. Therefore outflow and inflow galaxies seem to show different trends and outflow events seem to be linked to relatively high metallicity. Again, as discussed in Section \ref{subsec:MSscatter}, the latter result is not likely due to selection effects.

As for the MS, in order to quantify this latter result we used a KS-test to compare the distribution of the residuals of the outflow galaxies to the one of the control sample. We find a p-value of 0.09, corresponding to a confidence interval of 91\%. 
This result suggests that galaxies associated to an outflow event are preferentially more metal-rich.
\subsection{Regarding fluctuation timescales}
\label{subsection:timescales}
In this work, we are probing  gas flows on CGM-scales (at impact parameters of 20 to 80 kpc), i.e. on
time scales of several hundreds of Myr, given that $\sim 250$ Myr is the typical time it takes for a gas cloud to travel to an average distance of $50$ kpc at speeds of 200 km/s.
This means there might be a time delay between these gas flow events and the measured SFR and metallicity, where nebular lines are sensitive to fluctuations on a few Myr ($<10$ Myr). Thus, outflows (inflows) events could be related more to the past (future) SFR/metallicity, respectively than to the current ones. 

However, recent cosmological simulations  showed that the scatter of the MS and MZR scaling relations is actually dominated by long timescale fluctuations of $\sim 1-10$ Gyr  \citep{Tacchella2016,Matthee2019,Iyer2020,vanLoon2021}, associated with the dark matter halo formation time \citep{Zenocratti2022}.
Therefore, if galaxies  fluctuate around the main sequence and mass-metallicity relation on such long timescales, one should see the impact of these fluctuations on the CGM on timescales of a few hundreds of Myr, up to several Gyr. 

The results shown in this work suggest that galaxies have preferential locations on the MS and FMR scaling relations according to CGM gas flows events. Galaxies associated to gas inflow events seem to be located on the upper part of the MS envelop, while galaxies associated with outflow events are located on the upper part of the FMR residuals. Hence, these results tend to support the long time scales for gas flow fluctuations.

\section{Summary and Conclusions}

\label{sec:summary}
We study the potential link between gas in-/outflows events and the location of galaxies on three key scaling relations. We first derive these  scaling relations (main sequence, mass-metallicity relation, and the fundamental metallicity relation) from a sample of 370 galaxies in our MUSE data using optical line ratios and the metallicity calibration from \citet{Curti2020}.
Using the same calibrators and methods, we use MUSE and X-Shooter data
to measure the SFRs and metallicities of 21 galaxies with known active gas flows (7 inflows and 14 outflows) based on MgII absorption measurements in background quasars and an impact parameter $b < 70$~kpc (papers~II and III).
This ensures no systematic offset between the parent sample and our SFGs with gas flows.


Overall, with this sample of 21 galaxies with {\it simultaneous} measurements of SFR, $M_\star$, metallicities and gas in-/outflow rates, we find that:

\begin{itemize}
\item galaxies with large inflows events have elevated SFRs and thus are preferentially located above the main-sequence (Figure \ref{fig:MS}), in agreement with the expectations from bathtub/gas-regulator models  \citep[e.g][]{Bouche2010,Lilly} where accretion powers SFR;
\item galaxies with inflows are preferentially located on the lower part of the MZ relation (5 out of 7, Figure \ref{fig:FMR}, left), but follow the same FMR as the general population (Figure \ref{fig:FMR}, right). Given that such accreting material seems to be metal-poor \citep{Wendt2021}, gas accretion dilutes the metallicity and simultaneously boosts the SFR, again supporting gas-regulator models;
\item galaxies with outflows are preferentially more metal rich than galaxies with inflows, at a given stellar mass (Figure \ref{fig:FMR}, left) and
tend to be slightly more metal rich in the FMR residuals (Figure \ref{fig:FMR}, right).
\end{itemize}


The potential connection between the CGM properties and the galaxy properties implies that galaxies are evolving on scaling relations with fluctuations over long timescales (of several hundreds of Myr) on average.
The potential differences between inflows and outflows selected samples might  originate from the combination of the travel time delays of the gas (which is approximately $+250$ Myr for outflows and $-250$ Myr for inflows, totaling to a time difference of $\lesssim$ 500~Myr-1Gyr) and the varying cosmological accretion rates.
\\

Other key properties are necessary to understand the bathtub equilibrium scenario, such as the gas content which has also been found to correlate with the scatter of the scaling relations discussed in this paper (e.g, \citealt{Tacconi2018},  \citealt{Ginolfi2020b},  \citealt{Freundlich2021}). Therefore, observations of gas masses (from CO or dust continuum) are needed to improve our understanding of galaxy evolution. 

Finally, this work highlights the necessity of larger samples of galaxies with simultaneous gas-phase metallicities measurements and gas flows information. In order to understand the full picture, we need samples of galaxies for which we have all the main properties, i.e., stellar mass, SFR, metallicity and molecular gas. The Atacama Large Millimeter/submillimeter Array (ALMA) could help investigate the latter property. 
 Nonetheless, the presented results support the theoretical picture according to which accretion of fresh gas drives star-formation while diluting the metal content, whereas outflows occur in more evolved (or more metal-rich) galaxies. 




\section*{Acknowledgements}
This work has been carried out thanks to the support of the ANR 3DGasFlows (ANR-17-CE31-0017). We thank the referee for a constructive report which improved the clarity and quality of the manuscript. \\
This work made use of the following open source software: MPDAF (\cite{Piqueras2017}), matplotlib (\cite{Matplotlib}), NumPy (\cite{Oliphant2007}, \cite{vanderWalt2011}, \cite{Numpy}), Scipy (\cite{Scipy}), Astropy (\cite{Astropy2013}, \cite{Astropy2018}). 

\section*{Data availability}
The work presented in this paper is based on observations collected at the European Organisation for Astronomical Research in the Southern Hemisphere under ESO programmes: 094.A-0211(B), 095.A-0365(A), 096.A-0164(A), 097.A-0138(A), 098.A-0216(A), 099.A-0059(A), 0100.A-0089(A), 0100.A-0089(B), 0101.A-0287(A), 0102.A-0712(A),  0104.B-0655(A) (available on the ESO archive \url{http://archive.eso.org}). The data from this work will be shared on \url{https://megaflow.univ-lyon1.fr} or via reasonable requests to the corresponding author.




\bibliographystyle{mnras}
\bibliography{biblio.bib}


\appendix{Appendix}
\section{Examples of MUSE and X-Shooter spectra}

\begin{figure*}
    \begin{center}
        \includegraphics[width=2\columnwidth]{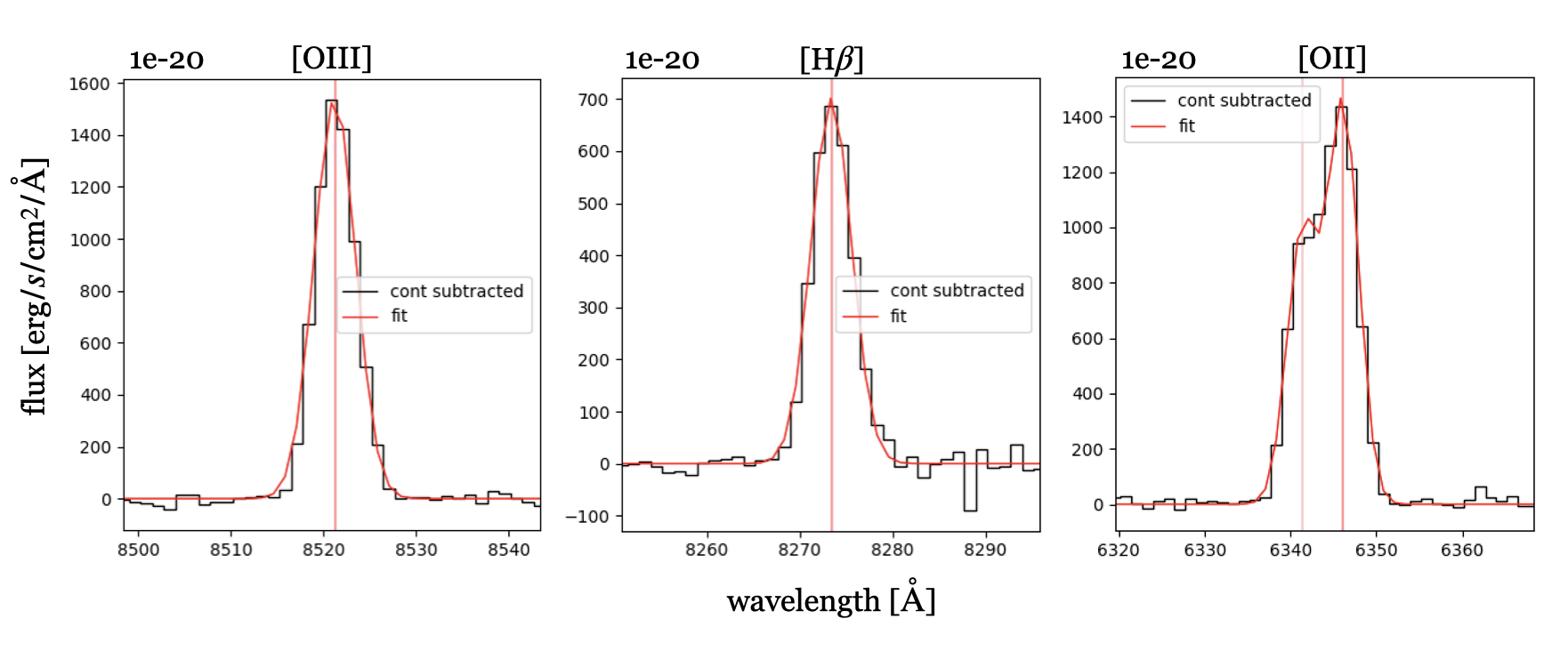}
        \caption{One high SNR example of MUSE spectra of the lines used in this work, namely, [{\rm O}{\sc \,iii}], $H\beta$ and [{\rm O}{\sc \,ii}] (galaxy 211130). The black line shows the continuum-subtracted spectrum, the red line shows the fit and the vertical red line shows the position of the peak, in all panels.}
        \label{fig:muselines}
    \end{center}
\end{figure*}

\begin{figure*}
    \begin{center}
        \includegraphics[width=1.8\columnwidth]{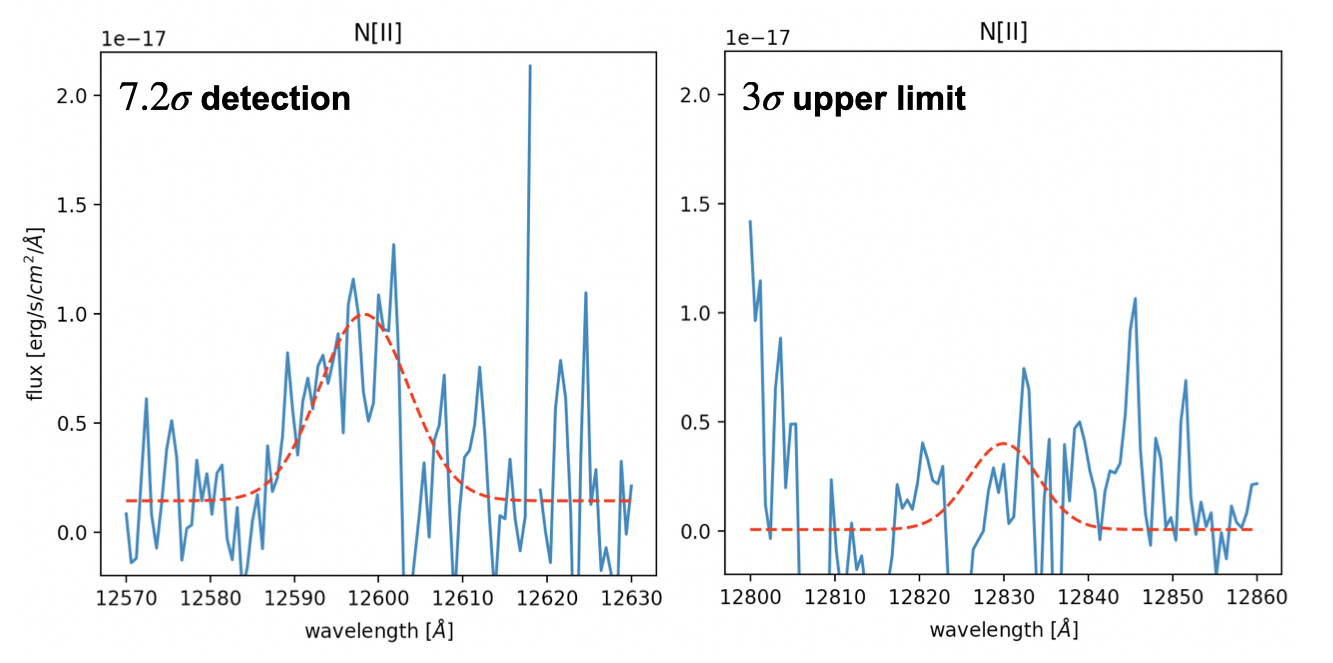}
        \caption{X-Shooter spectra of a robust [{\rm N}{\sc \,ii}] detection (galaxy 52 on the left) and a [{\rm N}{\sc \,ii}] $3 \sigma$ upper limit (galaxy 39 on the right). The red dashed line shows the gaussian fit from which we extract the integrated flux, where in the case of a $3 \sigma$ upper limit we set the peak of the gaussian to be 3 times above the noise limit and the width of the gaussian to be the same as the one fitted to the $H\alpha$ line.}
        \label{fig:detanduplim}
    \end{center}
\end{figure*}

In this first appendix, we show an example of the spectra and their best fits for the MUSE data (Figure \ref{fig:muselines}), as well as for the X-Shooter data (Figure \ref{fig:detanduplim}). In the case of X-Shooter data, we show one spectrum for which we have a robust detection and one for which we apply a $3 \sigma$ upper limit due to a combination of poor data and the difficulty of detecting the faint [{\rm N}{\sc \,ii}] line.

\section{Distribution of the properties of our sample and a comparison of the metallicity derived from MUSE and X-Shooter data}

\begin{figure*}
    \begin{center}
        \includegraphics[width=1.8\columnwidth]{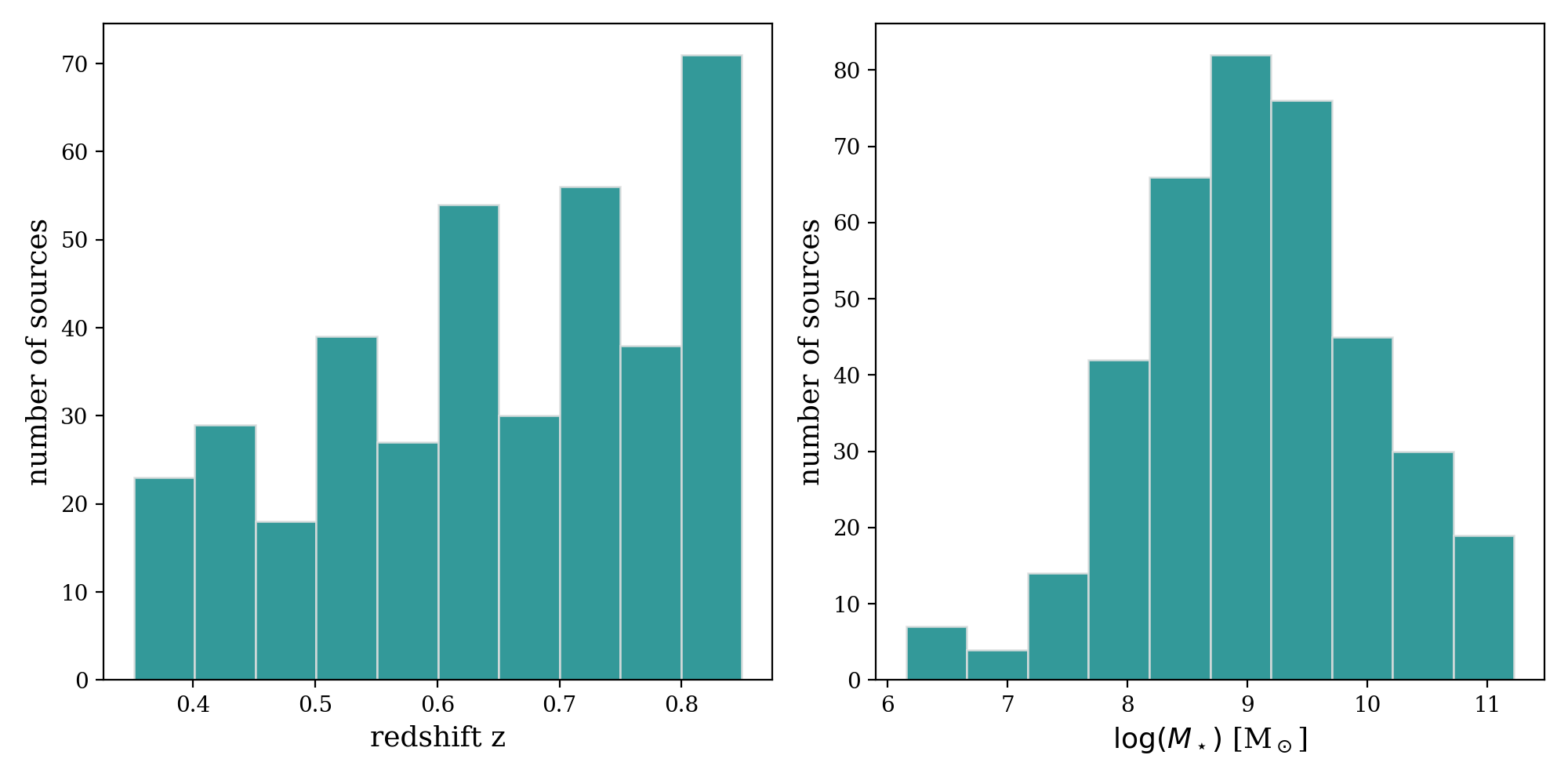}
        \caption{Left: Redshift distribution of our sample going from $z=0.35$ to $z=0.85$. Right: Stellar mass distribution of our sample going from $M_* = 10^{6.2} \, \text{M}_\odot$ to $M_* = 10^{11.3} \, \text{M}_\odot$. Each distribution adds up to 385 galaxies, the number of galaxies we have in our MUSE sample.}
        \label{fig:histograms}
    \end{center}
\end{figure*}

In this second appendix, we show the different properties of our MUSE sample (Figure \ref{fig:histograms}), namely the redshift and stellar mass distributions. Adding the 6 X-Shooter galaxies extends the redshift distribution to $z = 1.4$, given the very small number of galaxies compared to the MUSE sample, we decided not to include them in the histogram of the redshift distribution for a better visualisation. In addition, Table \ref{tab:xshooter_muse_comp} compares the metallicities derived from rest-optical emission lines with MUSE data and the metallicities derived from rest-optical emission lines redshifted in the NIR with X-Shooter data, for the same galaxies.

\begin{figure*}
    \begin{center}
        \includegraphics[width=1.8\columnwidth]{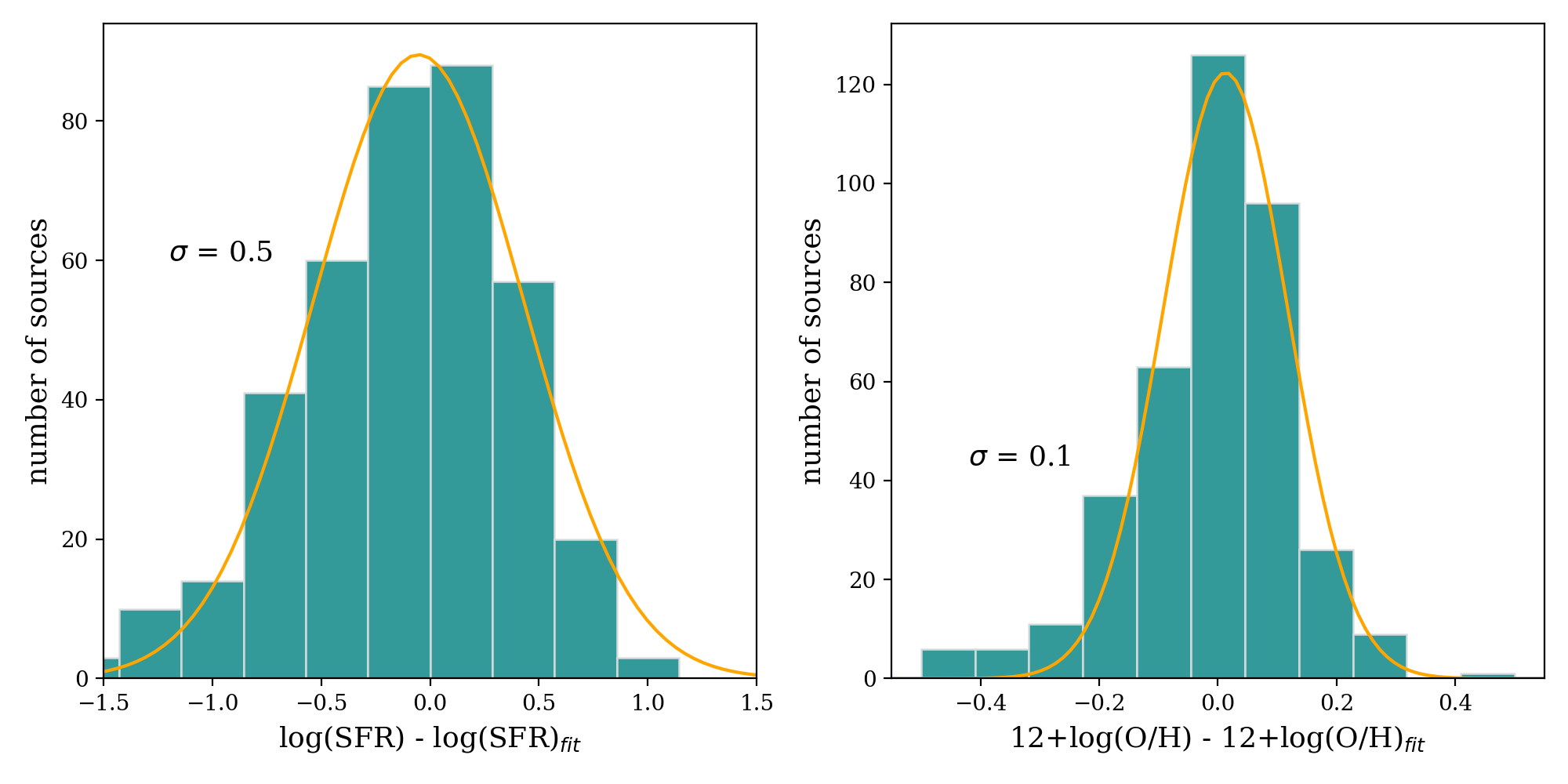}
        \caption{Left: The main sequence residuals distribution. Right: The fundamental metallicity relation residuals distribution. Both panels show the distribution for the entire sample minus the galaxies for which the error on the stellar mass is too large, $\sigma(M_\star) > 0.4 \log(\text{M}_\odot)$, as these galaxies do not have a significant impact on the fits due to their errors.}
        \label{fig:histograms_res}
    \end{center}
\end{figure*}

Figure \ref{fig:histograms_res} shows the residuals distribution for the main-sequence and the fundamental metallicity relation. After fitting a Gaussian to the residuals, without taking into account the galaxies for which the error on the mass is large ($\sigma(M_\star) > 0.4 \log(\text{M}_\odot)$, grey circles), we find a scatter of $\sigma \approx 0.5$ and $\sigma \approx 0.1$ for the MS and FMR respectively.

\begin{table*}
	\centering
	\caption{Comparison of the metallicity derived from MUSE, using the combination of three metallicity tracers ($R_3$, $R_{23}$, $O_{32}$), corresponding to the "$\text{Metallicity}_{\text{MUSE}}$" column and the metallicity derived from X-Shooter, using only one metallicity tracer ($N_2$), corresponding to the "$\text{Metallicity}_{\text{X-Shooter}}$" column.}
	\begin{tabularx}{2\columnwidth}{XrXXXXX}
	    \hline
		MUSE ID & X-Shooter ID & Metallicity$_{\text{MUSE}}$ & $\sigma_{\text{MUSE}}$ & Metallicity$_{\text{X-Shooter}}$ & $\sigma_{\text{X-Shooter}}$\\
		\hline
		\hline
		131065 & 91 & 8.74 & 0.05 & 8.74 & 0.05\\
		\hline
		192003 & 24 & 8.58 & 0.05 & 8.69 & 0.20\\
		\hline
		211087 & 33 & 8.58 & 0.05 & 8.60 & 0.05\\
		\hline
		152003 & 9 & 8.47 & 0.05 & 8.40 & 0.05\\
	\end{tabularx}
	\label{tab:xshooter_muse_comp}
\end{table*}

\bsp	
\label{lastpage}
\end{sloppypar}
\end{document}